\documentstyle[prl,twocolumn,aps]{revtex}
\input{epsf}
\begin{document}
\draft
\twocolumn[\hsize\textwidth\columnwidth\hsize\csname @twocolumnfalse\endcsname
\title{Cooper problem in the vicinity of Anderson transition}

\author{J. Lages and D. L. Shepelyansky$^{*}$}

\address {Laboratoire de Physique Quantique, UMR 5626 du CNRS, 
Universit\'e Paul Sabatier, F-31062 Toulouse Cedex 4, France}

\date{November 29, 1999}

\maketitle

\begin{abstract}
We study numerically the ground state properties of the Cooper 
problem in the three-dimensional Anderson model. It is shown that
attractive interaction creates localized pairs in the metallic 
noninteracting phase. This localization is destroyed at sufficiently
weak disorder. The phase diagram for the delocalization transition in
the presence of disorder and interaction is determined.
\end{abstract}
\pacs{PACS numbers:  74.20.-z, 72.15.Rn, 05.45.Mt}
\vskip1pc]

\narrowtext


The pioneering experimental results for normal-state resistivity of high
temperature superconductors demonstrated a striking correlation between
the optimal doping with maximal $T_c$ in the superconducting phase and 
the Anderson metal-insulator transition (MIT) in the normal phase obtained 
in a strong pulsed magnetic field \cite{greg}. More recent experiments
on the superconductor-insulator transition (SIT) in three dimensions (3D)
\cite{lavrov,gantmakher,ando}, which were done in various 
materials at different dopings and magnetic fields, also 
reveal close correlation between these transitions even
if it is possible that the normal state remains metallic
in some materials \cite{lavrov,ando}. These experimental results
put forward the important theoretical problem of interaction
effects in the vicinity of Anderson transition in 3D. However, the full
understanding of this problem is very difficult since
even the origin of the high-$T_c$ phase is not yet established
completely. Due to that it would be interesting to understand 
the effects of interaction and disorder in a more simple model
of generalized Cooper problem \cite{cooper} of 
two quasiparticles above the frozen Fermi
sea which interact via the attractive Hubbard interaction
in the presence of disorder. 
In spite of apparent simplicity of this problem it is
rather nontrivial. Indeed, even if the great progress has been 
reached recently in the investigation of localized one-particle eigenstate
properties \cite{mirlin}, the analytical expressions for the
interaction induced matrix elements in the localized phase and
in the MIT vicinity are still absent. Furthermore, the recent 
results for the problem of two interacting particles (TIP) in the 
localized phase demonstrated that the interaction effects for
excited states can qualitatively change the eigenstate structure
leading to the appearance of delocalization
\cite{ds94,imry,pichard,moriond}.
Due to that the investigation of the ground state properties 
of the above model in the vicinity of the Anderson transition 
in 3D represents an interesting unsolved problem which can 
shed light on the origin of SIT in the presence of disorder.

To investigate the above problem we study numerically the ground state
properties of  two particles with Hubbard on 
site attraction ($U<0$) in 3D Anderson model at half filling.
In this case the one particle eigenstates are determined by the 
Schr\"odinger equation 
\begin{equation}
\label{hamil}
E_{\mathbf n}\psi_{\mathbf n}+V (\psi_{\mathbf n-1}+\psi_{\mathbf n+1})
=E\psi_{\mathbf n}
\end{equation}

\begin{figure}
\epsfxsize=8.5cm
\epsfysize=10cm
\epsffile{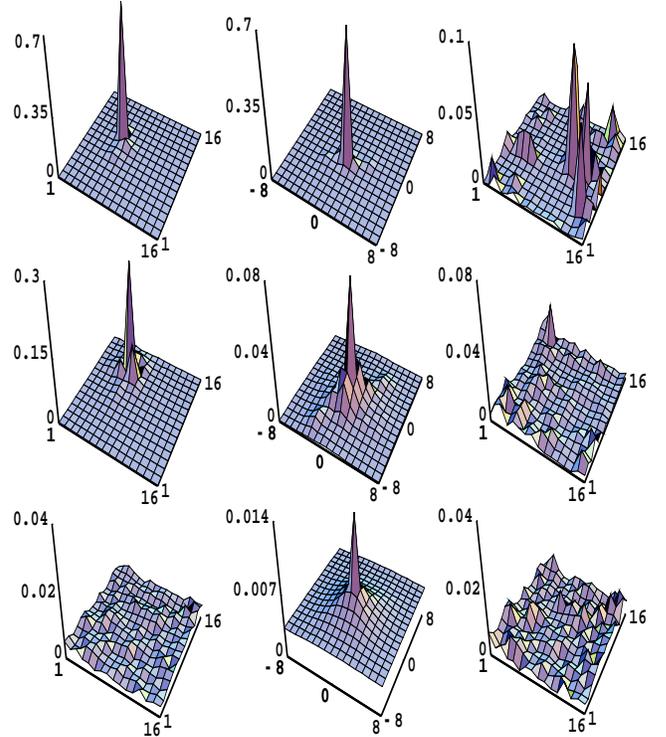}
\vglue 0.2cm
\caption{Probability distributions for TIP in the ground state projected
on $(x,y)$-plane: one particle probability $f_p$ for interaction
$U=-4 V$ (left column),
interparticle distance probability $f_{pd}$ for $U=-4 V$ (centrum
column), $f_p$ for $U=0$ (right column); the disorder strength is 
$W/W_c=1.1$ (upper line), $W/W_c=0.5$ (middle line), $W/W_c=0.3$ (bottom line).
All data are given for the same realisation of disorder for the system size
$L=16$ (see text for details). Upper line corresponds to the insulating 
noninteracting phase while two others are in the metallic one at $U=0$.} 
\label{fig1}
\end{figure}

\noindent where ${\mathbf n}$ is the site index on the 3D lattice
with periodic boundary
conditions applied, $V$ is the nearest 
neighbour hopping and the random on-site energies $E_i$ are homogeneously
distributed in the interval $[-W/2,W/2]$. It is well known that
at half filling (the band center with $E=0$) the MIT takes place
at $W_c/V \approx 16.5$ with the insulating and metallic phases
at $W>W_c$ and $W<W_c$ respectively (see e.g. \cite{shklovskii,japan}).
To study this problem with interaction 
it is convenient to write its Hamiltonian
in the basis of noninteracting eigenstates of the Anderson
model that gives 
\begin{eqnarray}
\label{tip}
(E_{m_1}+E_{m_2})\chi_{m_1, m_2} & + & 
U \sum_{{m^{'}_1}, {m^{'}_2}} Q_{m_1, m_2, {m^{'}_1}, {m^{'}_2}}
 \chi_{{m^{'}_1}, {m^{'}_2}} \nonumber \\
  & = & E\chi_{m_{1}, m_{2}}.
\end{eqnarray}
Here $\chi_{m_{1}, m_{2}}$ are eigenfunctions of the TIP problem
written in one-particle eigenbasis $\phi_m$ with eigenenergies $E_m$.
The transition matrix elements $Q_{m_1, m_2, {m^{'}_1}, {m^{'}_2}}$
are obtained by rewriting the Hubbard interaction in the
noninteracting eigenbasis of model (\ref{hamil}). 
The Fermi sea is introduced by restricting the sum in (\ref{tip}) to
$m^{'}_{1,2}>0$ with unperturbed energies $E_{m^{'}_{1,2}} > E_F$. The
value of the Fermi energy $E_F \approx 0$ is determined by the filling factor 
$\mu$ which is fixed at $\mu = 1/2$. To have more close similarity with the
Cooper problem we also introduce the high energy cut-off defined by 
the condition $1 \leq m^{'}_1+m^{'}_2 \leq M$.
Such a rule gives an effective phonon frequency 
$\omega_D \propto M/L^3$ where $L$ is the linear lattice size.
Since the frequency $\omega_D$ should be independent of $L$ we keep
the ratio $\alpha=L^3/M$ constant when varying $L$. The majority of
data are obtained for $\alpha \approx 30$ \cite{note} but we checked that its 
variation by few times did not affect the results. 
Due to on-site nature of the Hubbard interaction only symmetric
configurations are considered.

In fact the first studies of the model
(\ref{tip}) with the frozen Fermi sea had been done by Imry \cite{imry}
with the aim to analyze the delocalization effect of TIP in the proximity
of Fermi level at finite particle density. This model was also studied
numerically in \cite{jacquod} where it was shown that near $E_F$ 
the interaction becomes effectively stronger comparing to the ergodic
estimate used in \cite{ds94,imry}. However the above studies
\cite{imry,jacquod} were concentrated on the properties of excited states 
in the repulsive case $U>0$. On the contrary here we analyze the ground state 
properties for the attractive case. Since $U<0$, then
even in the limit of large system size $L$ 
the particles are always close to each other in the ground state
that is qualitatively different 
from the case $U>0$. In this way the model (\ref{tip}) represents
the generalized Cooper problem in the presence of disorder.

To study the characteristics of the ground state $\chi^{(0)}_{m_1, m_2}$ we 
diagonalize numerically the Hamiltonian (\ref{tip}) and rewrite the
eigenfunction in the original lattice basis $|{\mathbf n} \rangle$ with the
help of relation between lattice basis and one particle eigenstates
$|{\mathbf n} \rangle = {\sum_{m} R_{{\mathbf n},m}} \phi_m$. As the result
of this procedure we determine the two particle probability distribution
$F({\mathbf n_1,n_2})$ in the ground state (here ${\mathbf n_{1,2}}$ mark the
positions of the two particles), from which the one
particle probability $f({\mathbf n_1})=\sum_{{\mathbf n_2}}F({\mathbf n_1,n_2})$
and the probability of interparticle distance 
$f_d({\mathbf r})=\sum_{{\mathbf n_2}}F({\mathbf r+n_2,n_2})$ with 
${\mathbf r}={\mathbf n_1}-{\mathbf n_2}$ are extracted. 
For graphical presentation 
these probabilities are projected on $(x,y)$-plane that gives 
$f_p(n_x,n_y)=\sum_{n_z}f({n_x,n_y,n_z})$ and $f_{pd}(x,y)$ respectively.
The typical examples of projected probability distributions $f_p$ and $f_{pd}$
for different values of disorder $W$ are shown in Fig. 1. They clearly show 
that in the presence of interaction the ground state remains localized not only
in the noninteracting localized phase ($W>W_c$) but also in the phase 
delocalized at $U=0$ ($W<W_c$). However the localized interacting 
phase abruptly disappears if disorder $W$ becomes smaller than some 
critical value $W_s(U)<W_c$.
For $W<W_s$ the ground state becomes delocalized over the whole lattice. At the
same time the peaked structure of the interparticle 
distance distribution $f_{pd}$
clearly shows that the particle dynamics remains correlated. In this sense we
can say that the pairs exist for any strength of disorder but for $W>W_s$ they
are localized while for $W<W_s$ they become delocalized. We assume that
such a transition should correspond to the transition from insulating 
to superconducting phase in the many-body problem.

\begin{figure}
\epsfxsize=8cm
\epsfysize=8cm
\epsffile{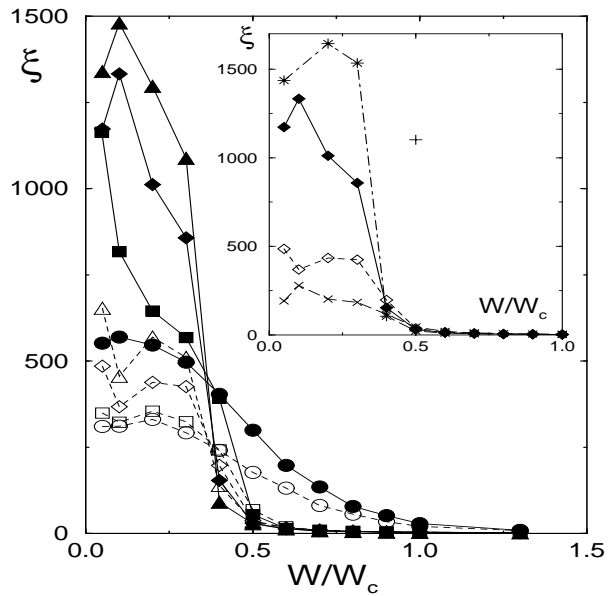}
\vglue 0.2cm
\caption{Dependence of IPR $\xi$ on the rescaled disorder strength $W/W_c$
for $U=0$ ($\circ$), $U=-2V$ ($\Box$), $U=-4V$ ($\diamond$), $U=-6V$
($\bigtriangleup$) (empty/full symbols are for $L=10 / L=12$). Insert shows 
the data for $U=-4V$ at $L=8$ ($\times$), $10\,(\diamond)$, $12$ (full
diamond), $14\,(*)$; $\xi$ obtained in the Cooper approximation (see text)
for $U=-4V$, $L=14$ is shown by (+). Statistical error-bars are smaller than
symbol size. 
Lines are drawn to adapt an eye.} 
\label{fig2}
\end{figure}

To analyse this transition in a more quantitative way we determine the inverse
participating ratio (IPR) $\xi$ for one particle probability :
$1/\xi=\langle \sum_{{\mathbf n}}f^2({\mathbf n}) \rangle$, where brackets mark
the averaging over 100 disorder realisations. Physically, $\xi$ counts the
number of sites occupied by one particle in the ground state. Its variation
with system size $L$ is shown in Fig. 2 for different strength of interaction
and disorder. This figure shows that in 
the localized interacting phase $W>W_s(U)$ the $\xi$
value remains finite and independent on size $L$ while in the delocalized phase
it grows proportionally to the total number of sites $L^3$. 
To find the critical
disorder strength $W_s$ we compare the relative change of $\xi$ with $L$
($8 \leq L \leq 14$) with its relative change for the noninteracting case at 
the critical point $W=W_c$. Then $W_s(U)$ is defined 
as such a disorder at which
the relative variation of $\xi$ at $|U|>0$ 
becomes larger than in the case $U=0$.
We note that near the transition the change of $\xi$ with $W$ is so sharp that 
the delocalization border is not really sensitive to the choice of definition.
We also checked that the change of $\omega_D$ does not affect significantly
the border $W_s(U)$ \cite{note1}. 
The phase diagram for SIT defined in the way described above is presented in
Fig. 3. It shows that the interaction makes localization stronger so that 
the localized phase penetrates in the noninteracting metallic phase. However
for sufficiently weak disorder delocalization takes over. Qualitatively we can
say that the attraction creates a pair 
with a total mass ($m_p$) twice larger than the
one particle mass and due to that the critical disorder strength becomes
twice smaller ($W_s/W_c \simeq 0.5$) since the effective hopping 
$V_{eff} \propto 1/m_p$. Of course this argument is not sufficient to explain
the exact border $W_s(U)$ obtained numerically but it gives a reasonable
estimate in the case of strong interaction. Further studies are required 
to explain the form of the border.

\begin{figure}
\epsfxsize=8cm
\epsfysize=8cm
\epsffile{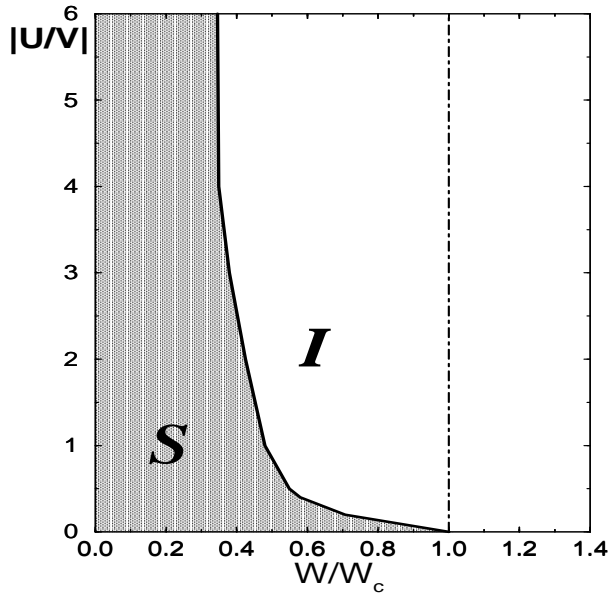}
\vglue 0.2cm
\caption{Phase diagram for transition from localized (insulating I) 
to delocalized (superconducting S) phase 
in the ground state of the generalized Cooper problem (\ref{tip}).
Vertical dashed line shows the Anderson transition in absence of interaction.} 
\label{fig3}
\end{figure}

Another interesting physical characteristic is the coupling energy $\Delta$
of two particles in the presence of interaction. Its value is equal to
$\Delta=2E_F-E_g$ where $E_g$ is the ground state energy in the presence of
interaction and $2E_F$ is equal to $E_g$ at $U=0$. In the standard Cooper 
problem $\Delta>0$ is related to the BCS gap and determines the correlation
length of the pair. It is interesting to understand how $\Delta$ varies
with the disorder strength $W$ at fixed interaction $U$. This dependence is 
presented in Fig. 4. It clearly shows that $\Delta$ grows significantly 
with the
\begin{figure}
\epsfxsize=8cm
\epsfysize=8cm
\epsffile{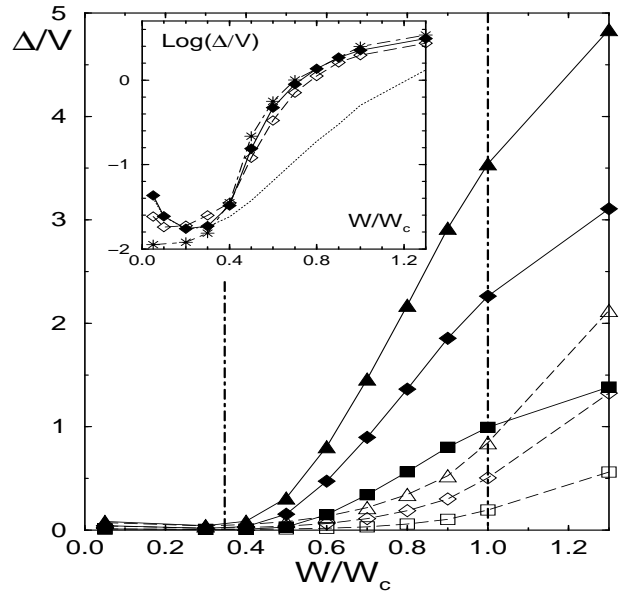}
\vglue 0.2cm
\caption{Variation of the ground state  coupling energy $\Delta$
in the model (\ref{tip})
with the rescaled disorder strength $W/W_c$ for $L=12$ and 
different interaction
$U=-2V$ (full box), $U=-4V$ (full diamond), $U=-6V$ (full triangle);
the open symbols are for the same $U$ values but in the Cooper 
approximation (see text). 
The vertical dashed line $W/W_c=1$ marks the MIT at $U=0$;
the other vertical dashed line $W/W_c=0.35$ marks approximately
the SIT line from Fig. 3 for $|U| \geq 2$.
Insert shows the case $U=-4V$ for $L=10$ 
($\diamond$), $L=12$ (full diamond), $L=14$ (*); dotted line shows 
the Cooper approximation case with $U=-4V$ and $L=12$ from the main 
figure; logarithm is decimal. } 
\label{fig4}
\end{figure}
\noindent increase of $W$ at {\it constant interaction} $U$. 
We attribute the physical origin of this growth
to the fact that at stronger disorder the rate of separation between particles
becomes smaller that enhances enormously the interaction between them,
hence $\Delta$, as it was discussed in \cite{jacquod}. 
The dependence of $\Delta$ on $W$ is changed drastically near
$W_s$ that is related to the delocalization transition.

On the same figure 
we compare the exact value of $\Delta$, found numerically in the model
(\ref{tip}), with its value $\Delta_C$ obtained by the Cooper approximation 
(mean field value). In this approximation only the matrix elements
$Q_{m_1, m_2, {m^{'}_1}, {m^{'}_2}}$ 
with $m_1=m_2$ and $m^{'}_1=m^{'}_2$ are kept in (\ref{tip}) that
corresponds to the original Cooper ansatz \cite{cooper}. The comparison
shows that at weak disorder $\Delta_C$ is very close to exact $\Delta$ 
(see insert where dotted line coincides with full diamonds 
for $W/W_c < 0.35$) while when approaching the Anderson transition and 
beyond it ($W/W_c > 0.35$) $\Delta_C$ becomes much smaller than $\Delta$.
This leads to the conclusion that the nondiagonal matrix elements
($m_1 \neq m_2$ and $m^{'}_1 \neq m^{'}_2$), neglected in the Cooper 
approximation, play an important role near MIT. This is also clear from
Fig.1 according to which the localized states exist in the noninteracting
metallic phase while according to the Cooper approximation pairs should
be delocalized for $W < W_c$. Indeed, for example the graphical image 
as in Fig. 1 shows that for $U=-4V$, $W=0.5W_c > W_s$ the probability $f_p$
obtained in the Cooper approximation from (\ref{tip}) is completely
delocalized contrary to the real case in which the ground state 
is localized (Fig. 1). In addition to this case the average IPR within
the Cooper approximation is much larger than its real value
obtained without approximation (see insert in Fig. 2).

The data presented in the inserts of Fig. 2 
and 3 clearly show that in the localized interacting phase $W>W_s$ 
the lattice size is sufficiently larger than the localization length
and the values of $\xi$ and $\Delta$ correspond to the limit 
$L \rightarrow \infty$ \cite{note2}.  
At the same time for weak disorder $W/W_c < 0.2$
the asymptotic value of $\Delta$ is very small and very large values 
of $L$ are required to reach it. Such large $L$ are also desirable
to see better the propagation of pairs with large size.
The main increase of $\Delta$ takes
place in the metallic noninteracting phase at $W_s < W < W_c$. However 
in this region the TIP pair remains localized due to interaction that
does not allow to obtain a gain in the value of $\Delta \simeq T_c $.
It would be interesting to find some possibility to delocalize the 
pair in this region and to keep large $\Delta$ at the same time. 
 
In conclusion, our numerical studies of the generalized Cooper problem
in the presence of disorder show that 
in the ground state the attractive interaction 
leads to localization of pairs inside 
noninteracting metallic phase, contrary to the Cooper ansatz.
This localization however disappears 
at sufficiently weak disorder. The phase diagram for the transition 
to delocalized states is determined as a function of disorder 
and interaction.

We thank V.V.Flambaum, K.Frahm and O.P.Sushkov for stimulating discussions,
and the IDRIS in Orsay and the CICT in Toulouse for access to 
their supercomputers.


\begin{thebibliography}{99}
\bibitem{note} In this case we found that 
               $\omega_D/V \approx 0.5 W/W_c+ 0.1$.
\end{thebibliography}
\end{document}